\documentclass[twoside]{dis07}
\usepackage[latin1]{inputenc}
\usepackage[dvips]{graphicx,epsfig,color}
\usepackage{wrapfig,rotating}
\usepackage{amssymb,amsmath,array}

\begin{document}
\title{A Review of Recent Results from the Tevatron}

\author{Giorgio Chiarelli~\cite{url}
%
\thanks{for the CDF and D0 Collaborations}
%
\vspace{.3cm}\\
%
INFN Sezione di Pisa \\
Largo B.Pontecorvo 3, I-56127, Pisa - Italy
%
}

\maketitle

\begin{abstract}
The D0 and CDF experiments have been taking data at the Run 2 of the 
Tevatron Collider since 2001.
We present a selection of recent results, most of them obtained with an integrated
luminosity of $\simeq$ 1 fb$^{-1}$.
I will describe the most important facets of the physics programme and 
detail some results. Recent direct limits on standard model Higgs
obtained at the Tevatron, and their
their prospects will be also reviewed.
\end{abstract}

\section{Introduction}
The D0 and CDF experiments are two $4\pi$ multi-purpose detectors taking data at 
the Tevatron Collider. 
Run 2 (started in 2001) is designed to provide each detector 
with $4-8$ fb$^{-1}$ of $p\bar{p}$ collisions by the end of 2009.
With respect to Tevatron Run I, the accelerator complex underwent a 
large upgrade which radically changed the way it operates. The 
interbunch distance was reduced from 3.5$\mu s$ to 392 ns, the whole 
$\bar{p}$ production, cooling and stacking was revised. As a result, after 
a relatively long startup, the accelerator is now performing very well. The peak luminosity
reached 2.92 10$^{32}$cm$^{-2}$s$^{-1}$ and is now delivering about 40 
pb$^{-1}$/week with a record of 45 pb$^{-1}$ in a single week.
Based on the current performances, the integrated luminosity per 
experiment extrapolates to $6-8$ fb$^{-1}$/experiment by the end of 
2009.

The CDF and D0 detectors were upgraded fully to exploit the physics
opportunities provided by the Tevatron. CDF completely rebuilt its tracking system (both 
the outer chamber and the silicon tracker), the forward calorimeter, 
its trigger and front end electronics and 
extended the muon coverage.
It also added the capability to trigger on tracks at Level1 (i.e. syncronous 
with the bunch
crossing) and to identify and trigger on tracks displaced with respect to the primary
vertex at Level 2. The silicon tracker is an important asset of its 
physics programme with
a precision single sided layer located right on the beam pipe, five layers of double sided
silicon sensors at various radii between 2.5 and 10 cm and two layers located at
$\simeq$ 20 and $\simeq$ 28 cm covering $|\eta|<2$ and $1<|\eta|<2$ 
respectively.

D0 changed its philosophy by becoming a full magnetic spectrometer with the addition of
a 2 Tesla superconducting solenoid. It also replaced its old tracker with a 
new 8-layer
fiber tracker which -combined with a microvertex silicon detector- provides a powerful
instrument to reconstruct tracks coming from the primary vertex and  
offline to identify 
vertices due to long-lived particles. D0 also improved its acceptance for muons and upgraded
the trigger system. Recently, in the shutdown of 2006, the collaboration added an extra 
layer of silicon sensors located right outside the beam pipe. 
 
The detectors collect data with an efficiency of $\simeq$ 90 \%. 
The small inefficiency is partly due to a deadtime coming from the trigger 
and Data Acquisition and partly to operational constraints. As we write 
$\simeq 2.5$ fb$^{-1}$ were written to tape by each experiment. However,
in the following, unless otherwise indicated, I will present results  
obtained with $\simeq 1$ fb$^{-1}$, less than half of the data on tape. 

\section{Flavour Physics}
Despite the large production cross section, 
processes involving HF remain largely buried under a large background.
CDF Run 1 pioneered the identification of heavy flavour at hadron 
colliders by detection of secondary vertices, a powerful tool 
complementary to other tagging techniques. 
In Run 2 the experiment
increased its B-physics reach by adding the capability to 
trigger on tracks not coming from the primary vertex (SVT). A number of 
$b$ and $c$ -related physics 
processes, otherwise completely buried by a large background, can 
theferore be selected online for further analyses.
Thanks to its microvertex detector and to its large muon coverage D0 is 
also able to perform a number of measurements.
Details can be found in ~\cite{bph}, here I only mention a few, very 
interesting results.

\begin{wrapfigure}{r}{0.5\columnwidth}
\centerline{\includegraphics[width=0.45\columnwidth]{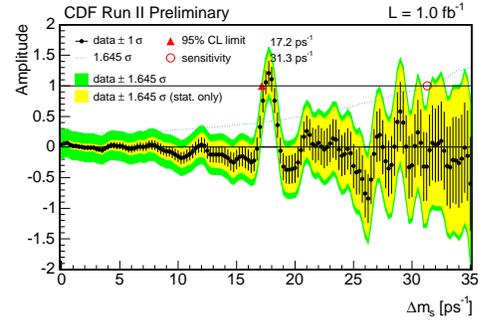}}
\caption{Amplitude scan for $\Delta m_s$ at CDF.}
\label{Fig:bosc}
\end{wrapfigure}

In Spring 2005 D0 presented a limit on $B_s$ oscillations using 0.9 
fb$^{-1}$ of $14.9<\Delta m_s< 21$ ps$^{-1}$ at 90\% C.L. With 1 
fb$^{-1}$ CDF presented (Fall 2005) a 5 $\sigma$ observation of $B_s$ 
oscillations (Fig.~\ref{Fig:bosc}) and a measurement of $\Delta 
m_s=17.77\pm 0.10(stat)\pm0.07(syst)$ ps$^{-1}$.
D0 exploits a combination of its measurement of the $B_s\rightarrow J/\Psi 
\phi$ channel and of the $B_s$ semileptonic decays together with 
results from the
$B$ factories and CDF $\Delta m_s$ to obtain a measurement of 
$\phi_s=0.70^{+0.47}_{-0.39}$~\cite{bph}.

CDF and D0 search for rare B decays. Thanks to SVT, CDF 
directly measures $B\rightarrow hh$ decays and their $A_{CP}$. 
$B_d$,$B_s$ decays to $\mu \mu$ have tiny SM branching fractions 
(O(10$^{-9}$) which are enhanced (by powers of $\tan \beta$) 
in several SUSY models, therefore both Collaborations search 
for new physics through this channel. CDF has not yet updated its 
measurement performed with 0.8 fb$^{-1}$, while D0 just presented its 
result with the full dataset of 2 fb$^{-1}$. Combining the 2a (without 
the silicon layer on the beampipe) and 2b 
data they find 3 candidate events with a background of 2.3$\pm$ 0.7 and 
set 
a limit for $B_s\rightarrow \mu \mu$ $<9.3(7.5)\cdot 10^{-8}$ at 95(90) \% 
C.L. and $B_d\rightarrow \mu \mu$ $<2.3(2.0)\cdot 10^{-8}$ at 95(90) \% 
C.L. This result (which will soon be improved by adding the CDF search),
sets interesting limits on many SUSY models by excluding zones in the 
$\tan \beta-M_A$ plane~\cite{carena}.
We expect that by 2009, with 8 fb$^{-1}$, the Tevatron will be able to set 
a limit of $\approx 2 \cdot 10^{-8}$ on the $B_s$ decay. 
\begin{wraptable}{l}{0.5\columnwidth}
\centerline{\begin{tabular}{|l|r|}
\hline
state& Mass value $\pm$ stat. $\pm$ syst.\\\hline
$\Sigma_B^+$ & $5808^{+2.0}_{-2.3}\pm1.7$\\\hline
$\Sigma_B^-$ & $5816^{+1.0}_{-1.0}\pm1.7$\\\hline
$\Sigma_B^{*+}$ & $5829^{+1.6}_{-1.8}\pm1.7$\\\hline
$\Sigma_B^{*-}$ & $5379^{+2.1}_{-1.9}\pm1.7$\\
\hline
\end{tabular}}
\caption{$\Sigma_B$ and $\sigma_B^*$ masses (in MeV/c$^2$).}
\label{tab:sigmab}
\end{wraptable}

Searches for rare decays of known states are complemented by the search 
for new states, and CDF recently presented the observation of two new $B$ 
baryons: $\Sigma_B$ and $\Sigma_B^*$ with masses shown in 
table~\ref{tab:sigmab}.
Last but not least, new measurements of $\Lambda_B$ lifetime are presented. 
CDF measures (exclusive 
states) $1.5\pm 0.077\pm 0.012$ ps while D0 reports $1.28\pm 0.11\pm 0.09$ 
ps in semileptonic decays and $1.3\pm 0.14\pm 0.05$ ps in 
exclusive channels~\cite{bph}.

\section{QCD and jet physics}
Tests of the strong interaction and measurements of jet distributions 
have 
been the bread and butter physics at the Tevatron for more than 20 years.
Besides testing theoretical prediction those processes are often used to 
test algorithms but they also play an important role to estimate the 
background in rare processes and searches for new physics.
This dual aspect is present in many analyses. Details can be found in the
many Tevatron contributions to this 
Conference~\cite{jet}. In the following I will only refer to a small 
subset.

 \begin{wrapfigure}{l}{0.5\columnwidth}
\centerline{\includegraphics[width=0.4\columnwidth]{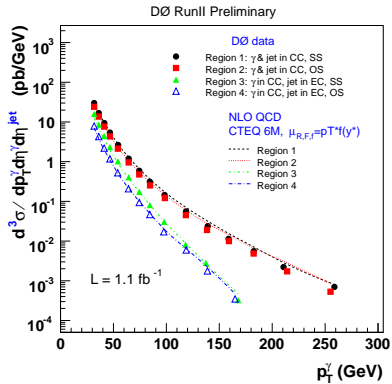}}
\caption{D0 triple differential $\gamma-j$ cross section. }
\label{Fig:d0gj}
\end{wrapfigure}

Thanks to the large statistics available it is now possible to 
make precisios measurements of 
associated production of jets and vector bosons ($W$, $Z$, $\gamma$).  
D0 measures the triple differential cross section in the $\gamma-jet$ 
process (Fig.~\ref{Fig:d0gj}) where finds a good agreement with 
available NLO QCD calculations.
CDF finds a good agreement between its inclusive jet distribution and 
theoretical prediction. The difference between theory and CDF data of Run 
1, has now disappeared.
 \begin{wrapfigure}{r}{0.5\columnwidth}
\centerline{\includegraphics[width=0.38\columnwidth]{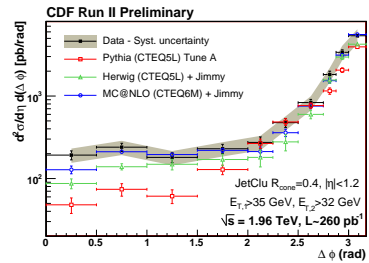}}
\caption{$\Delta \varphi$ between two identified $b$s }
\label{Fig:cdfbb}
\end{wrapfigure} 
The large statistics, combined with the exploitation of the SVT trigger.
allows CDF to study $b\bar{b}$ correlations in a small dataset ($\simeq
260$ pb$^{-1}$).
Existing MC do not fully describe the data, as you can see from 
Figure~\ref{Fig:cdfbb} where several MC are used for the comparison.
The region in which the two $b$s are close in $\varphi$ exhibits a clear
deviation of data with respect to calculations. Even the improvement 
obtained by using Jimmy (a Monte Carlo describing multiple parton 
interactions) does not fully account for this difference, which happens in 
the region where gluon splitting is expected to provide a sizeable 
contribution to the process.

\section{Electroweak Physics Measurements} 
Electroweak (EWK) processes 
can be used to understand better the capabilities of the detectors and to 
develop new tools (from trigger to analysis technique). They also often 
represent background for searches. 
With 
its large dataset, the Tevatron became a place where precision EWK 
measurements can be performed to test the SM at its boundaries.

\begin{wrapfigure}{r}{0.5\columnwidth}
\centerline{\includegraphics[width=0.45\columnwidth]{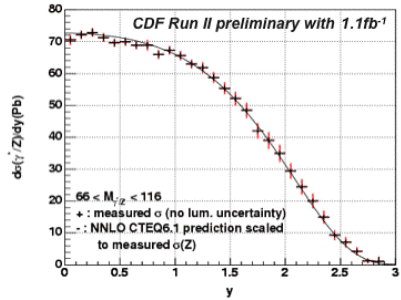}}
\caption{$d\sigma/dy$ for $Z\gamma*\rightarrow ee$ CDF events.}
\label{Fig:dsdy}
\end{wrapfigure}

Among the many results, I chose a few which are significant for their 
implications.
The $W$ and $Z$ integral and differential cross sections provide an excellent 
testing ground for
PDFs. NLO and NNLO calculations of the inclusive processes have been available for quite some 
time and recently a full differential calculation at NNLO became 
available~\cite{melnikov}. The large 
statistics collected allows CDF to present a $d\sigma/dy$ for 
$Z\rightarrow ll$ events (Figure~\ref{Fig:dsdy}).
While the agreement with theory is good, this measurement -with 
increasing statistics- can be
used to costrain PDFs. 
Recently CDF measured the ratio of central-to-forward cross section for 
$p\bar{p}\rightarrow W+X$ with the $W\rightarrow e \nu$ and demonstrated a 
sensitivity of this quantity to PDFs. This can be a promising way to study
PDFs at the LHC where the $W$ asymmetry measurement will play a less 
prominent role~\cite{wplug}.

Tevatron experiments recently obtained significant results in the
diboson sector. The tiny cross sections  (of the order of a few pb, 
less than the top cross section) challenged
the experimentalists' determination and ingenuity. 
Diboson production represents also 
a test bed for the detection of Higgs and 
indeed represents a background in several channels.

In Summer 2006 D0 presented its evidence of $WZ$ production, with $WZ 
\rightarrow l\nu l\nu$  
signal at 3.3 $\sigma$ level and a 
cross section of $3.98^{+1.91}_{-1.53}$ pb (statistical and systematic 
uncertainty combined), consistent with
SM expectation of $3.7\pm 0.1$ pb.
In Winter 2007 CDF
confirmed its previous evidence observing a signal of 16 events with 
a background of $2.7\pm 0.28\pm0.33\pm 0.09$. The probability of a null
signal is $<1.5 \times 10^{-7}$ equivalent to a $\simeq 6 
\sigma$ effect. The measured
cross section is $5.0^{+1.8}_{-1.6}$ pb 
(statistical and systematic uncertainty combined). 
In this case the improvement did not come from a larger dataset 
but rather from improved analysis technique and a larger lepton 
acceptance.

\begin{wrapfigure}{l}{0.5\columnwidth}
\centerline{\includegraphics[width=0.4\columnwidth]{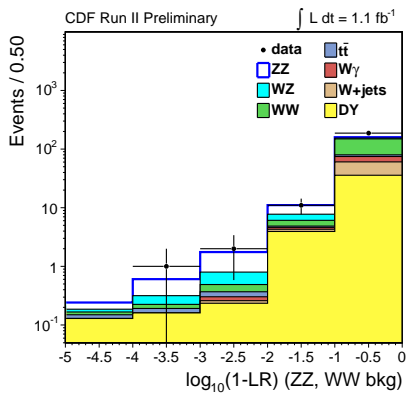}}
\caption{Log-Likelihood ratio for $ZZ$ candidates superimposed to
background.}
\label{Fig:zz}
\end{wrapfigure}

Another challenging process is $ZZ\rightarrow llll$. CDF shows a 
3$ \sigma$ evidence (Figure~\ref{Fig:zz})
and measures a cross section of $0.75^{+0.71}_{-0.54}$ pb, 
compatible with the
NLO prediction of $1.4\pm 0.1$ pb.
At present the case in which one of the two bosons decays hadronically 
still remains unobserved.

\begin{wrapfigure}{r}{0.5\columnwidth}
\centerline{\includegraphics[width=0.45\columnwidth]{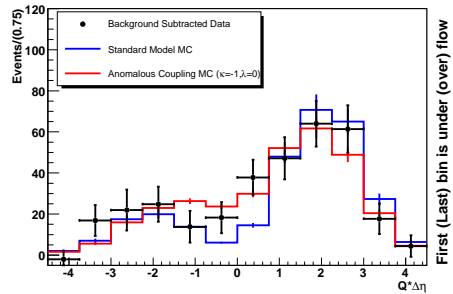}}
\caption{Signed $\Delta \eta_{l,\gamma}$ distribution at D0.}
\label{Fig:wgamma}
\end{wrapfigure}

The gauge structure of the SM finds a crucial test in the associated production of
$W\gamma$. The destructive interference at tree level 
of the relevant diagrams 
creates a zero in the $dN/d\cos\theta^*$ distribution at $cos\theta^*=\pm 
{1\over{3}}$, where $\theta^*$ 
is the c.o.m. angle between the $W$ and the incoming quarks.
In our detectors we measure the charged lepton from the $W$ decay and the
sensitive variable is $Q\cdot\Delta \eta_{l,\gamma}$ where Q is charge of 
the lepton. The distribution of this quantuty still shows a 
dip at $\approx -0.3$.
As the photon does not directly couples to the $Z$ the interference is not
present in the $Z\gamma$ process.
Both experiments measure the
inclusive $W\gamma$ and $Z\gamma$ production cross section. CDF finds
$\sigma(W\gamma) = 19.1\pm 2.8$ pb and $\sigma(Z\gamma)=4.9\pm 0.5$ pb. D0 
applies a cut to
the photon $E_T$ ($>$7 GeV) and to the transverse mass 
M($\gamma$,$l$,$\nu$) $> 90$ GeV and
quotes $\sigma(W\gamma)=3.2\pm 0.5\pm 0.2$(lum) and $\sigma(Z\gamma)$ of 
$4.51\pm 0.4 \pm 0.3$ pb.
D0 measures the $Q\cdot \Delta \eta$ distribution  
in 900 pb$^{-1}$ and in its data there is evidence of a dip related to the destructive intereference 
predicted by the SM (Figure~\ref{Fig:wgamma}).

In the single boson realm, the most significant contribution 
came from CDF which directly measured the $W$ mass and width.
The traditional way is to study the transverse mass 
($M_T=\sqrt{2 \cdot E^{\nu}_T \cdot E^l_T \cdot(1-cos \theta_{l,\nu})}$)
distribution where $l=e$,$\mu$ and neutrino transverse momentum is
estimated from the transverse missing energy size and 
direction. The peak
provides information about the $W$ mass where the non-Gaussian tail 
(due to the Lorentz distribution) about the
$W$ width. In its $M_W$ measurement CDF also fits the transverse momentum 
distributions of the leptons. 
\begin{wrapfigure}{l}{0.5\columnwidth}
\centerline{\includegraphics[width=0.35\columnwidth]{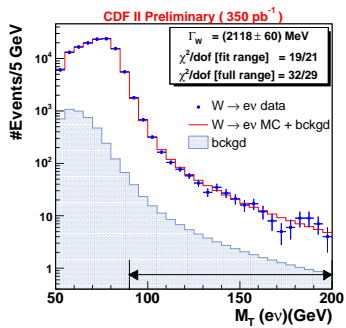}}
\caption{Fit to $M_T$ in the $W\rightarrow e\nu$ channel.}
\label{Fig:gammaw}
\end{wrapfigure}
The result is $M_{W}=80413\pm 48$ MeV/c$^2$ where statistical and 
systematic uncertainties
contribute evenly (34 MeV/c$^2$ each). This is the best measurement 
obtained by a single 
experiment. As it was performed on 200 pb$^{-1}$, while the 
current sample on tape exceeds it by a factor 12, it 
is reasonable to expect a large
reduction of the statistical error. The systematics 
can also be addressed with a larger statistics and a precision of $\simeq 
25$ MeV/c$^{2}$ seems achievable.
The measurement of $\Gamma_W =2032\pm 71$ MeV/c$^2$ (with 350 pb$^{-1}$)
(Fig.~\ref{Fig:gammaw}), 
combined with Run 1 measurements, now dominates the world average.

\section{Top Physics}
Top quark production was first discovered at the Tevatron Collider in 
1994-1995. 
Until the LHC starts it is still the only place where it can be studied 
and several new results are presented~\cite{top}. 

\begin{wrapfigure}{r}{0.5\columnwidth}
\centerline{\includegraphics[height=0.4\columnwidth]{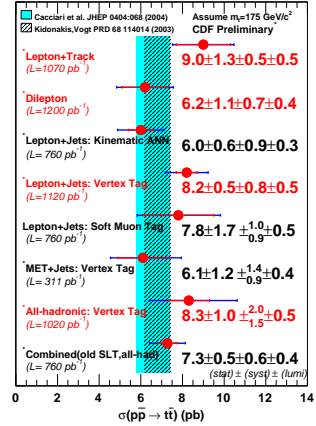}}
\caption{$t\bar{t}$ production cross section at CDF.}
\label{Fig:cdfxt}
\end{wrapfigure}

As the $t \rightarrow bW$
$\simeq$100 \% of the times, one can classify the various decay channels according to the way
the $W$ boson decays. In this way one can measure {\it dilepton} channel (both $W$s decay into
$l\nu$), {\it lepton+jets} channel where only one of the two $W$s decays leptonically and
finally the {\it all-hadronic} channel in which both $W$s decay hadronically. The final
states contain, accordingly, one, two or no high $P_T$ lepton.
The structure of the $Wtb$ vertex can be directly studied in top-pair decays.
Anomalous couplings (FCNC) and new physics might appear as deviation from SM expectations.
The Top cross section has been measured in essentially all decay channels. 
A compilation of the $t\bar{t}$ production cross section measured by 
CDF can be found in Fig.~\ref{Fig:cdfxt}.

Some comments are in order.
The dilepton channel, by far the one with least background has a BF of 
only $\simeq 4.9$ \% summing together the $ee$,$\mu \mu$,$e\mu$ channels. 
In order
to improve statistics CDF selects events with one fully identified 
high-$P_T$ lepton and
an isolated high $P_T$ track. Its recent result with 1.1 fb$^{-1}$ is 
$\sigma_{dil}=9.0\pm 1.3(stat)\pm 0.5(sys)\pm 0.5(lum)$ pb. 
D0 has a comparable result with a similar data sample: 
$\sigma_{dil}=6.8^{+1.2}_{-1.1}(stat)^{+0.9}_{-0.8}(sys)\pm 0.4(lum)$ pb. 

The $l+jets$ channel has worst signal-to background ratio, therefore, 
since
the beginning of top physics, it has been customary to exploit the 
presence of two jets containing $b$ quark. 
The characteristic signature due to the presence of long-lived
particles is used to improve S/B. One can require one or two $b$ tags 
(i.e. jets
identified as containing $b$ debris) with an efficiency that, for $t\bar{t}$ events
reaches $\simeq$ 55 \%. While both CDF and D0 are trying to improve their 
$b$ tagging algorithms
to increase efficiency, the overall acceptance and S/B ratio are already 
good enough 
to ensure that this channel is the most important in many top quark 
physics measurements.
As for the cross section, the most recent result comes from D0 (1 
fb$^{-1}$) and is
$\sigma_{l+j}=8.3^{+0.6}_{-0.5}(stat)^{+0.9}_{-1.0}(sys)\pm 0.5(lum)$ pb.
 
In the fully hadronic channel the final state (6 jets) has a large 
multi-jet background to compete with. Therefore, after triggering the
S/B is $\simeq 1/1300$. A combined neural-net based kinematic analysis 
and $b$ tagging improve this ratio to $\simeq 1/16$.
Thanks to this selection the result for the cross section is comparable
to the other two channels.
In $\simeq 1$ fb$^{-1}$ CDF measures:
$\sigma_{all-had}=8.3\pm 1.0(stat)^{+2.0}_{-1.5}(sys)\pm 0.5(lum)$ pb

With less than 50\% of the dataset analyzed, the cross section measurements
are reaching the level of the theoretical NLO calculations.
$\sigma_{t\bar{t}}=6.7^{+0.7}_{-0.9}$ pb~\cite{cacciari}, 
$\sigma_{t\bar{t}}=6.8\pm 0.6$ pb~\cite{kidonakis}.
A NNLO calculation might become quite interesting, 
even more when the LHC comes into operation although 
at the moment it appears too challenging to be addressed with standard 
calculation procedures.

Top decays before hadronizing, therefore there are no 
bound states,
unlike the other quarks. Therefore its mass, a fundamental quantity that 
combined with the $W$
mass, provides us on insight on the Higgs sector, can be accurately 
measured.
CDF and D0 measure $M_{top}$ in each decay channel using 
several techniques. The original {\it template} 
method, where distributions from data were compared with expectations from 
(combined) top MC and background, 
is now complemented by Matrix Element (ME) and Dynamic Likelihood Method (DLM)
where the intrinsic structure of the decay enters directly and helps to 
improve the measurement.

In table~\ref{tab:topmass} we summarize the
most recent results obtained with $\approx$ 1 fb$^{-1}$.
\begin{wraptable}{l}{0.5\columnwidth}
\centerline{\begin{tabular}{|l|r|}
\hline
All-hadronic (CDF, 943 pb$^{-1}$)  & $171.1 \pm 4.3$ \\\hline
Dilepton (CDF, 1030 pb$^{-1}$) & $164.5 \pm 5.6$ \\\hline
Dilepton (D0, 1000 pb$^{-1}$)  & $172.5 \pm 8.0$ \\\hline
Lepton+jets (CDF, 940 pb$^{-1}$)  & $170.9 \pm 2.5$ \\\hline
Lepton+jets (D0, 900 pb$^{-1}$)  & $170.5 \pm 2.7$ \\\hline
World Average            & $170.9 \pm 1.8$ \\
\hline
\end{tabular}}
\caption{Best $M_{top}$results (in GeV/c$^2$).}
\label{tab:topmass}
\end{wraptable}

The new world average is
$M_{top}=170.9 \pm 1.8$ GeV/c$^2$. While this measurement is largely dominated
(about 70 \%) by the results in the $l+jets$ channel, the all-hadronic channel
is acquiring a more prominent role. The systematic uncertainty in this 
measurement is already 2.1 GeV/c$^2$ close to the 1.4 
GeV/c$^2$ of the lepton+jets channel.

In previous measurements the dominant systematic effect came from the
Jet Energy Scale (JES). JES indicates 
all the effects that -for a given measured jet energy- provide us with the 
information about the energy of the original parton. Both experiments are 
now 
calibrating {\it in situ} the JES by exploiting the constraint provided 
by the jets coming from the hadronic $W$ decay. In this way the JES is 
included in the statistical uncertainty of the measurements and will 
improve with larger data sets. For example the statistical 
uncertainty of the CDF $l+jets$ measurement (166 events) includes
two contributions: 1.6 
GeV/c$^2$ from statistics and the remaining from JES.

\begin{wrapfigure}{r}{0.5\columnwidth}
\centerline{\includegraphics[width=0.45\columnwidth]{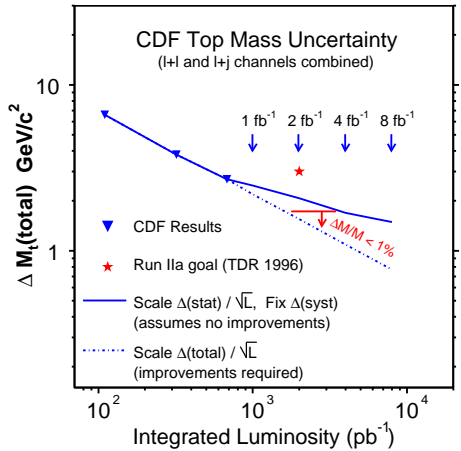}}
\caption{Prediction for $M_{top}$ accuracy as 
a function of integrated luminosity.}\label{Fig:persp}
\end{wrapfigure}
With more than 2 fb$^{-1}$ on tape, the future of top quark measurements 
looks bright.
The top mass will improve with the larger dataset, as the JES will be better 
constrained. Also, both experiments isolated a sample of $Z\rightarrow b\bar{b}$ events
that can be used to set the $b$ specific jet energy scale.
In figure~\ref{Fig:persp} we show the prediction for the top mass 
measurement at CDF. 
While, for example CDF is already doing
better than predicted in the TDR~\cite{CDFTDR}, it is difficult to 
establish 
what the asymptotic limit will be, but a precision $<$ 1 \% can be 
reached and the Tevatron can aim for a combined accuracy of $\leq$ 1 
GeV/c$^2$, making this measurement a long lasting Tevatron legacy.
Such an accuracy is, however, inducing both Collaborations to start 
addressing a number of effects that, too small to have an impact in the 
first measurements, can now become relevant.
Moreover, a more general discussion of the meaning of the quantity 
measured is 
in order. CDF and D0, use Pythia Monte Carlo to generate top 
templates to which they compare data. With such an accuracy, of the order 
of the top natural width, one should be careful in interpreting the 
meaning of the measurement, in particular as we make larger use of DLM and 
ME methods.

\begin{wrapfigure}{l}{0.5\columnwidth}
\centerline{\includegraphics[width=0.4\columnwidth]{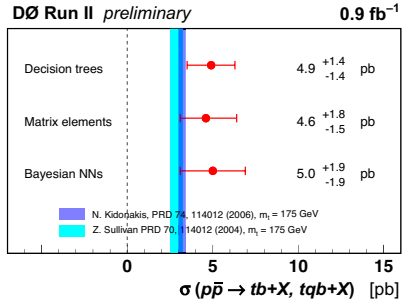}}
\caption{Evidence for single top.}\label{Fig:singlet}
\end{wrapfigure}

The most significant recent result in terms of $top$ production came from 
D0 which, for the first time, presented evidence for {\it single top} 
production.
This purely EWK process proceeds through two ($s$ and $t$) channels, which 
have SM cross section
of 0.88 and 1.98 pb respectively. While CDF sets a combined upper limit of 
2.6 pb at 95 \% C.L., D0 finds a 3.4 $\sigma$ signal
combining three analyses and advanced statistical 
techniques. 
The measured production cross section is $4.9 \pm 1.4 $pb 
(Figure~\ref{Fig:singlet}), as it 
is direcly proportional to $|V_{tb}|^2$, D0 is
able to set a direct limit $0.68<|V_{tb}|<1$ at 95 \% C.L.\\

\section{Higgs Searches}
\begin{wrapfigure}{r}{0.5\columnwidth}
\centerline{\includegraphics[height=0.33\columnwidth]{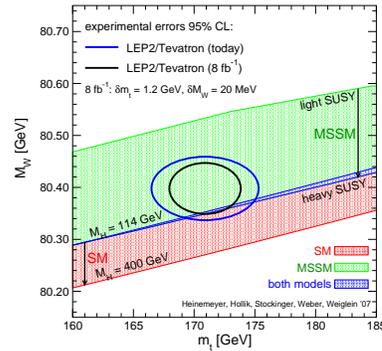}}
\caption{Constraints on SM and non-SM Higgs from 
indirect measurements}\label{Fig:higgs}
\end{wrapfigure}
In the SM the Higgs mass is directly connected to the $W$ and top mass,
therefore precision measurements of the $W$ and top mass mentioned above
translate into a limit $M_{H}<144$ GeV/c$^2$ at 95 \% C.L. which
rises to 182 GeV/c$^2$ if one takes into account the (direct) LEP 2 limit
of $M_{H}>114$ GeV/c$^2$. This result might imply some tension 
between
the SM prediction and the observation, however it only appears at
1 $\sigma$ level. 
Figure~\ref{Fig:higgs} shows the 95 \% C.L. contour which demonstrates how 
only by the end of Run 2 one might really gather (indirect) information 
on the Higgs SUSY sector from the $M_{top}$ and $M_W$ measurements.

While the indirect measurement was always seen as the major 
contribution of the Tevatron to Higgs hunting, in recent years the
increased luminosity delivered by the accelerator pushed the
two Collaborations aggressively to pursue direct Higgs searches.
The experimental situation at the Tevatron has two bounds. One is the 
cross section. For low mass Higgs ($<$ 120 GeV/c$^2$) direct 
production from gluon fusion is still $\leq$ 1 pb, while associated 
production of Higgs with $W$ or $Z$ boson is about an order of magnitude 
smaller.
\begin{wrapfigure}{l}{0.5\columnwidth}
\centerline{\includegraphics[height=0.25\columnwidth]{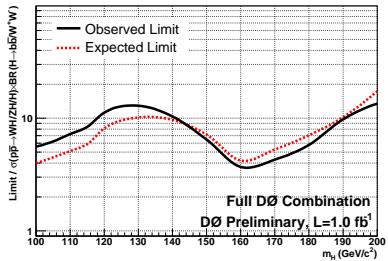}}
\caption{D0 direct Higgs limits($\simeq$ 1 fb$^{-1}$).}
\label{Fig:higgsdl}
\end{wrapfigure}
In this region Higgs decay $\approx$ 80\% of the time directly into 
$b\bar{b}$ pairs.
The huge background due to heavy flavour jets prevents us from searching 
in the $b\bar{b}$ channel while the low cross section prevents us from 
searching for this production mode through its rare -but almost background 
free- decays (like $\gamma \gamma$).
Therefore in this region we concentrate on the search in the $WH$ and $ZH$ channel
where the $W$ and $Z$ provides a clean signature and (most) of the triggering opportunities.
The Higgs decay into two $b$s can be exploited further to reduce the 
background by exploiting
the $b$-tagging technique, as already done in $top$ events.

Recently the large data sample available opened up the opportunity to
look for high mass Higgs ($\simeq 160$ GeV/c$^2$) directly produced by 
$gg$ fusion and decaying into $W$ pairs. 
By exploiting the leptonic decays of the $W$ the background is very low
and mostly due to SM processes. By increasing the acceptance as 
much as possible Tevatron experiments have become quite competitive.

Both D0 and CDF present results with 1 fb$^{-1}$ in several channels.
Figure~\ref{Fig:higgsdl} shows the combination of D0 results from many
channels across the whole mass range of searches. The ratio 95\% CL/SM is 
8.4 for $M_{H}=115$ GeV/c$^2$ and 3.7 for $M_{H}=160$ GeV/c$^2$.
With respect to Summer 2006 more analyses were performed and new 
techniques were used.

CDF has not yet provided a full combination of its searches. The most 
recent results, all with $\simeq$ 1 fb$^{-1}$ are in the  $ZH$,$Z\rightarrow 
ll$, $Z\rightarrow \nu \nu$ channels and in the $H\rightarrow WW^*$ 
channel. In the first two searches the ratio with respect to the SM cross 
section is 16 (for $M_H=115$ GeV/c$^2$) while for the third is 5.6 
for  $M_H=160$ GeV/c$^2$ (equivalent to a cross section limit of 2.2 pb).

Unfortunately no 
official Tevatron combined limit is yet available and, indeed, for 
example, the D0 combined limit alone is already better than the previous 
(Summer 06) Tevatron combined.
Despite that, it is clear that, even before the end of 
Run 2, the search for the Higgs at the Tevatron will provide useful 
input to the LHC experiments.
\begin{wrapfigure}{r}{0.5\columnwidth}
\centerline{\includegraphics[height=0.5\columnwidth]{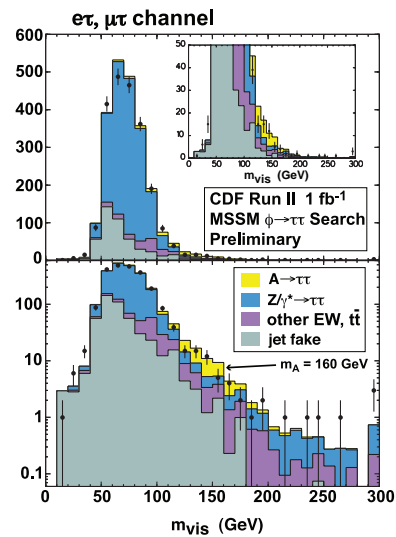}}
\caption{CDF Search for $H\rightarrow \tau \tau$. }\label{Fig:htt}
\end{wrapfigure}

\section{Searches for New Physics}
The Tevatron is not only performing precision measurements of the SM, but 
is testing 
its frontier to check a number of theories which have been proposed as well as
for any unknown possibility. It is not possible to fully present the whole
set of analyses, ranging from SUSY to Extra Dimensions, Leptoquarks and 
more, which are, however, discussed in other contributions to this 
Conference~\cite{bsm}. Therefore I will only present a sample of recent 
results.

The SUSY paradigm is intensively tested, as already discussed 
in the flavour sector.
First of all both CDF and D0 search for non-SM Higgs. As the SUSY Higgs 
has a large decay rate in
$\tau$ pairs, and its production can be enhanced for large $\tan 
\beta$
both experiments developed a number of $\tau$-ID algorithms to exploit
the good S/B ratio of the $H\rightarrow \tau \tau$ channel. 
Dedicated triggers and
extensive improvement of algorithms brought up the efficiency for this channel.
$\tau$ are identified through the detection of their debris in the $\tau 
\rightarrow l+\nu \nu$ and the $\tau \rightarrow$ hadronic decay.
CDF searched for a discrepancy from SM expectations in the visible mass
($m_{vis}$) 
distribution, where by $m_{vis}$ we mean the invariant mass of the visible 
$\tau$ decays and the missing $E_T$.
In the region $\simeq 150$ GeV/c$^2$ a 
small excess of
$\simeq 2 \sigma$ is visible in the channel where one of the two taus decays 
hadronically.
Figure~\ref{Fig:htt} shows a hypothetical Higgs with
mass $M_A=150$ GeV/c$^2$ superimposed on data.
In the corresponding channel in which both $\tau$s decay leptonically the search does not
have enough statistics to see a possible signal.
D0 result does not exhibit a similar enhancement.
\begin{wrapfigure}{r}{0.5\columnwidth}
\centerline{\includegraphics[width=0.45\columnwidth]{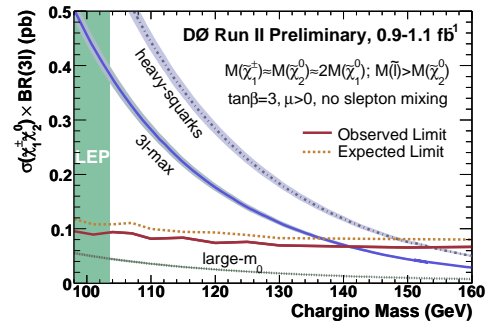}}
\caption{D0 limit for chargino production.}
\label{Fig:d0cn}
\end{wrapfigure}

Another SUSY sector being tested is through the direct search for
chargino and neutralino which are produced with 
sizeable cross sections. No signal is observed and therefore limits are 
set for $M_{\chi^\pm}$ (figure~\ref{Fig:d0cn}).
 

Despite the large background and the small cross section, CDF performed a 
search for direct squark and gluino production in a 
sample of events containing large missing transverse energy 
and three jets.
\begin{wrapfigure}{r}{0.5\columnwidth}
\centerline{\includegraphics[width=0.4\columnwidth]{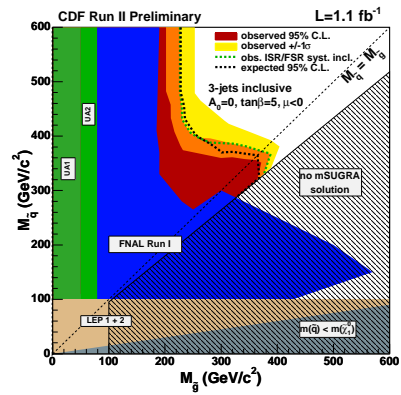}}
\caption{CDF limits in the $M_{Squark}-M_{gluino}$ plane.}
\label{Fig:cdfsg}
\end{wrapfigure}
The negative result is converted in a limit in the $M_{squark}-M_{gluino}$ 
plane (Fig.~\ref{Fig:cdfsg}).

\section{Conclusion}
With more than 2 fb$^{-1}$ already on tape, and the prospects of 
integrating between 6 and 8 fb$^{-1}$, the Tevatron experiments are now
testing the standard model at its boundaries. 
The detectors are well understood and the analyses are now mature, 
therefore the precision study of known processes can be used to measure
structure functions, test theoretical calculations and challenge 
measurements performed elsewhere.
The measurement of $M_W$ and $M_{top}$ can represent an enduring legacy of 
the Tevatron well after the LHC starts taking data.
The large datasets allow
to search for new physics and for the yet undetected Higgs particle which
now appears -for some mass ranges- within reach.

\section*{Acknowdledgments}
I would like to thank many CDF and D0 colleagues who helped me in 
preparing my contribution to this Conference. In particular my thanks to 
O.~Atramentov,  
P.~Bussey,
J.~Cammin, 
M.~Corcoran, 
M.~D'Onofrio, 
R.~Erbacher, 
S.~Jabeen, 
R.~Kehoe, 
M.~Heck, 
J.~Hobbs, 
S.~Leone,
C.~Mesropian, 
M.~Neubauer, 
L.~Pinera,  
B.~Reisert, 
A.~Robson,
P.~Strohemer, 
D.~Stuart, 
S.~Vallecorsa.
R.~Van Kooten, 
R.~Vilar,
J.~Wagner.
All mistakes are, of course, mine.


\begin{footnotesize}




\begin{thebibliography}{99}
\bibitem{higgs} Tevatron results for direct Higgs searches were presented 
in this conference by R.~Vilar.
\bibitem{bsm} Many searches for physics beyond the standard model are
discussed in the contributions to these Proceedings by 
P.~Strohemer and D.~Stuart,
\end{thebibliography}
%

\end{footnotesize}


\end{document}